\documentclass[conference]{IEEEtran}
\IEEEoverridecommandlockouts
\usepackage{cite}
\usepackage{amsmath,amssymb,amsfonts}
\usepackage{algorithmic}
\usepackage{graphicx}
\usepackage{textcomp}
\usepackage{xcolor}
\usepackage{subfig}
\def\BibTeX{{\rm B\kern-.05em{\sc i\kern-.025em b}\kern-.08em
    T\kern-.1667em\lower.7ex\hbox{E}\kern-.125emX}}
\begin{document}

\title{ Perceptual Quality Assessment for Video Frame Interpolation
\thanks{This work was supported in part by the National Natural Science Foundation of China under Grant 62271312, and in part by the Shanghai Pujiang Program under Grant 22PJ1407400.
Thanks for the support provided by OpenI Community (https://openi.pcl.ac.cn)}
}
\author{
\IEEEauthorblockN{1\textsuperscript{st} Jinliang Han, 2\textsuperscript{nd} Xiongkuo Min,
3\textsuperscript{rd} Yixuan Gao,
4\textsuperscript{th} Jun Jia,
8\textsuperscript{th} Guangtao Zhai}
\IEEEauthorblockA{\textit{Institute of Image Communication and Network Engineering} \\
\textit{Shanghai Jiao Tong University}, 
China \\
hanjinliang@sjtu.edu.cn, minxiongkuo@sjtu.edu.cn,
gaoyixuan@sjtu.edu.cn, jiajun0302@sjtu.edu.cn,\\zhaiguangtao@sjtu.edu.cn}
\IEEEauthorblockN{5\textsuperscript{th} Lei Sun, 6\textsuperscript{th} Zuowei Cao,
7\textsuperscript{th} Yonglin Luo}
\IEEEauthorblockA{\textit{Tencent}, 
China \\
raylsun@tencent.com, ernestcao@outlook.com,
luoylin2007@126.com}
}

\maketitle

\begin{abstract}
The quality of frames is significant for both research and application of video frame interpolation (VFI). 
In recent VFI studies, the methods of full-reference image quality assessment have generally been used to evaluate the quality of VFI frames. 
However, high frame rate reference videos, necessities for the full-reference methods, are difficult to obtain in most applications of VFI. 
To evaluate the quality of VFI frames without reference videos, a no-reference perceptual quality assessment method is proposed in this paper.
This method is more compatible with VFI application and the evaluation scores from it are consistent with human subjective opinions.
A new quality assessment dataset for VFI was constructed through subjective experiments firstly, to assess the opinion scores of interpolated frames. 
The dataset was created from triplets of frames extracted from high-quality videos using 9 state-of-the-art VFI algorithms. 
The proposed method evaluates the perceptual coherence of frames incorporating the original pair of VFI inputs. 
Specifically, the method applies a triplet network architecture, including three parallel feature pipelines, to extract the deep perceptual features of the interpolated frame as well as the original pair of frames. 
Coherence similarities of the two-way parallel features are jointly calculated and optimized as a perceptual metric. 
In the experiments, both full-reference and no-reference quality assessment methods were tested on the new quality dataset. 
The results show that the proposed method achieves the best performance among all compared quality assessment methods on the dataset.

\end{abstract}


\section{Introduction}
\begin{figure}[tb]
\centerline{\includegraphics[width=1.0\linewidth]{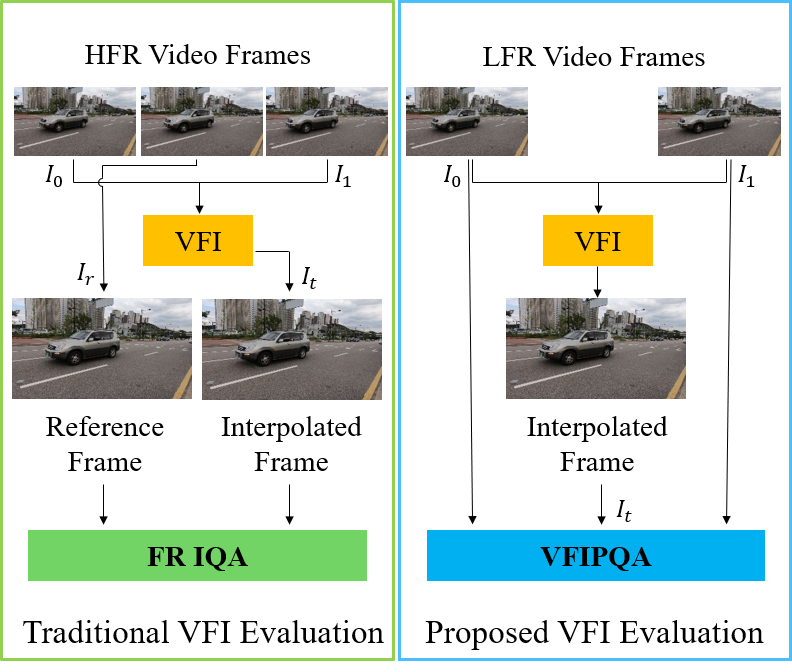}}
\caption{Comparison between traditional VFI evaluation and the proposed VFI evaluation frameworks.}
\label{fig0}
\end{figure}

Video frame interpolation (VFI) has been extensively studied and applied in computer vision and multimedia fields.
Its objective is to generate intermediate frames between original video frames.
The resulting video exhibits smoothness and increased frame rate, making it visually appealing and suitable for slow-motion playback\cite{jiang2018super}.
The VFI users pursue high-quality videos or even every frames produced by VFI algorithms. 
To evaluate the quality and compare performance of VFI algorithms, a typical evaluation framework is depicted in the left portion of Fig.~\ref{fig0}.
This framework employs full-reference (FR) image quality assessment (IQA) methods, commonly including Peak Signal-to-Noise Ratio (PSNR), Structural Similarity Index Measure (SSIM)\cite{wang2004image}, and Learned Perceptual Image Patch Similarity (LPIPS)\cite{zhang2018unreasonable}, which necessitate the extraction of reference frames from high frame rate (HFR) videos\cite{niklaus2017video, jiang2018super, bao2019depth, park2020bmbc, choi2020channel, liu2020enhanced, niklaus2020softmax, ding2021cdfi, huang2020rife}.

Applying FR methods to VFI encounters numerous challenges. While the traditional framework can be applied to evaluate VFI algorithms by generating a low frame rate (LFR) video for interpolation from a high frame rate (HFR) video, it may not be suitable for practical VFI applications where only LFR videos are available.
The FR approaches are restricted to applications in scenarios where the reference is available. 
A no-reference (NR) assessment approach appears to be more useful as it does not require any reference frame information \cite{zhai2020perceptual,8241862,10.1145/3470970}.
Moreover, certain VFI methods may introduce artifacts and blurred regions, especially when dealing with large moving objects or scene transitions between frames. The special distortions are noticeable to viewers, leading to a poor visual experience. But traditional metrics struggle to capture these frame interpolation distortions\cite{yang2008new}. In recent years, deep learning algorithms that extract high-level features of images have achieved significant success and demonstrated strong correlation with human perception\cite{zhang2018unreasonable, ding2020image}. The learning-based methods typically employ networks to extract deep quality features from the distorted images like humans and derive final image quality scores through training\cite{bosse2017deep, zhang2018blind, su2020blindly, yang2022maniqa}. However, none of learning-based methods have been specifically trained for VFI-specific image distortions. Therefore, a learning-based method for VFI perceptual quality assessment (VFIPQA) is proposed, aiming to address these gaps.

\begin{figure*}[htb]
\subfloat[]{\includegraphics[width=0.36\textwidth]{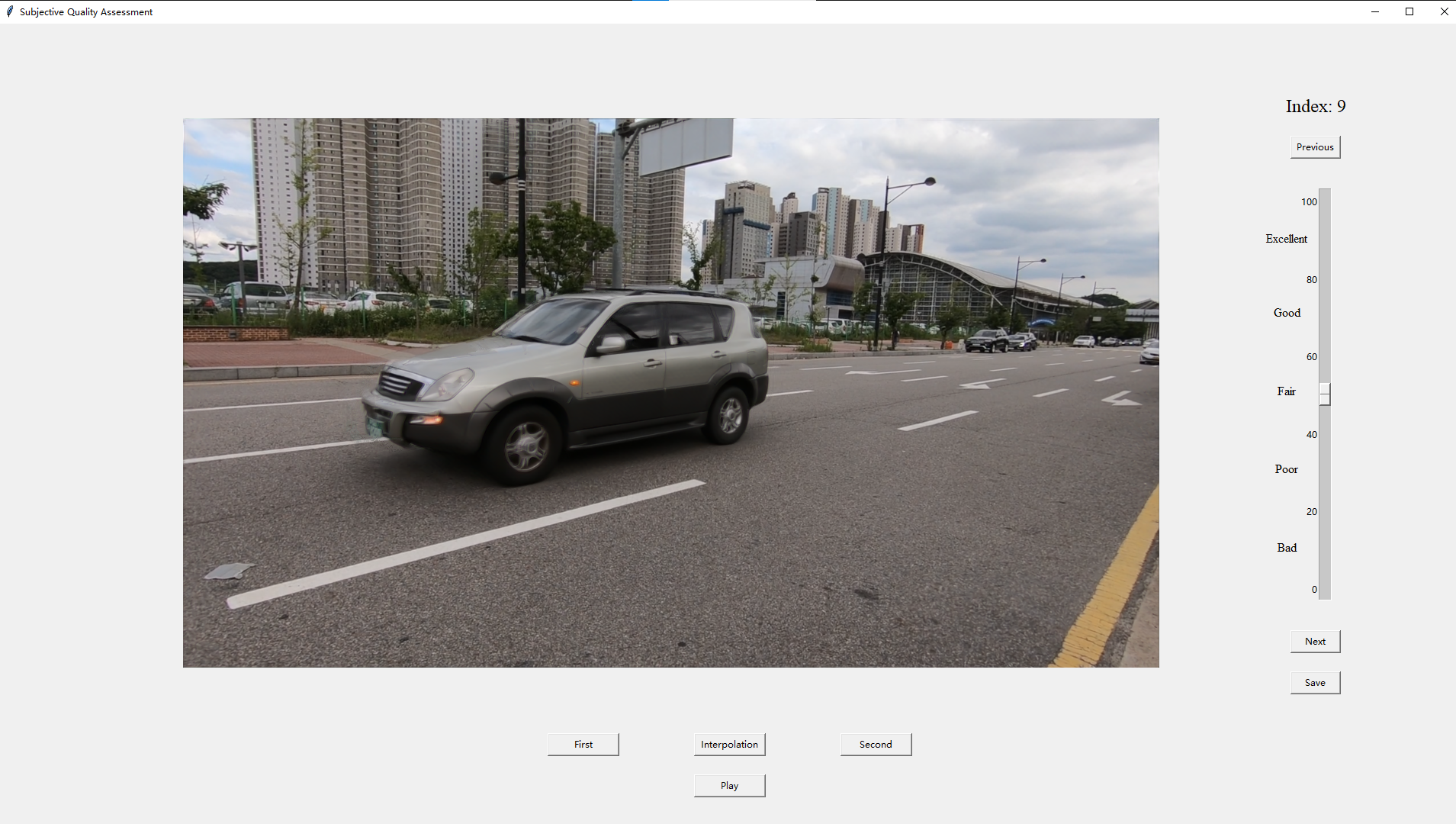}\label{fig02a}}
\hfill 
\subfloat[]{\includegraphics[width=0.3\textwidth]{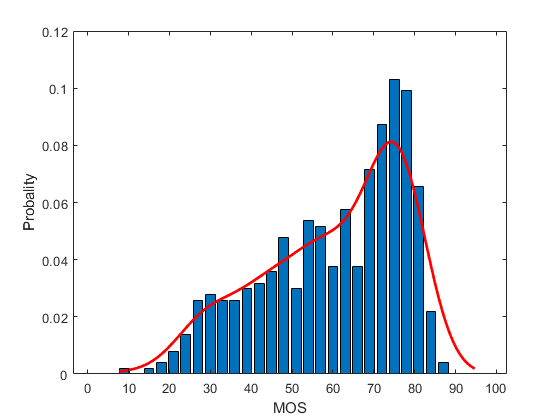}\label{fig02b}}
\hfill 
\subfloat[]{\includegraphics[width=0.3\textwidth]{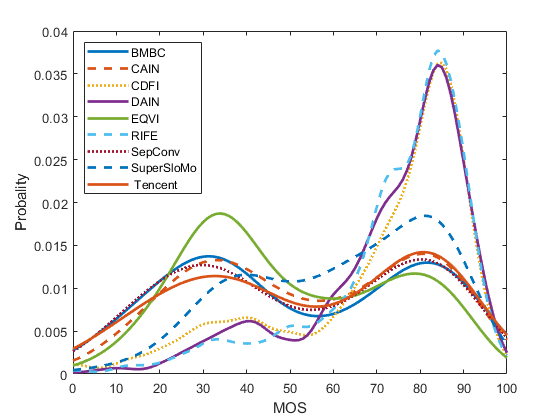}\label{fig02c}}

\caption{The design of the subjective experiment and the distribution of the experimental results. (a) is a screenshot of the designed GUI for subjective rating. (b) is MOS histograms and the fitted kernel distributions of the whole dataset. (c) is the fitted kernel distributions of MOSs for different VFI algorithms. }
\label{fig02}
\end{figure*}

This paper makes several follow contributions. Firstly, a novel VFI quality assessment dataset is developed. The dataset focuses on the quality assessment of single-frame interpolated from VFI algorithms.
Next, an NR perceptual quality assessment method is designed as depicted in the right part of Fig.~\ref{fig0}. 
Different from traditional NR IQA, the method takes both the interpolated frame and the original frame pair as inputs, which simulates the perceptual evaluation on frames coherence.
Deep features are extracted from frames and quality scores are computed by the designed coherence similarities between feature maps. 
Experimental results demonstrate that the proposed method outperforms on the dataset with the perceptual quality of VFI.

\section{Subjective Experiment and Dataset}

To the best of our knowledge, there is a lack of dedicated datasets for evaluating the quality of frames from VFI algorithms within the IQA communities.
In this section, we construct a new quality assessment dataset for interpolated frames.
The dataset is constructed using 56 original frame triples selected from a training dataset of VFI challenge\cite{Son_2020_ECCV_Workshops_VTSR}. This dataset consists of video clips spanning frame rates from 15fps to 60fps, providing a diverse range of scenarios for VFI. 
The 30fps video set, which is closer to the frame rate of most videos, was chosen as the source of frames for our dataset construction. Two consecutive frames from each video were randomly selected for interpolation, and the corresponding triplet frames in the 60fps set was chosen as the ground truth reference for FR methods.

To incorporate various types and levels of interpolated frame distortions produced by VFI algorithms, the distortion portion of the new dataset was generated using popular VFI algorithms. Eight algorithms from academia were selected, namely SepConv\cite{niklaus2017video}, SuperSloMo\cite{jiang2018super}, DAIN\cite{bao2019depth}, BMBC\cite{park2020bmbc}, CAIN\cite{choi2020channel}, EQVI\cite{liu2020enhanced}, RIFE\cite{huang2020rife}, CDFI\cite{ding2021cdfi}, and one algorithm developed by Tencent in practical applications. These algorithms are known to produce typical distortions in VFI. All the interpolated frames generated by algorithms were set with a frame rate up-scaling factor of two. After filtering out frames that failed to generate intermediate frames, the dataset consists of a total of 488 interpolated frames. Additionally, there are reference triplet frames containing 56 interpolated frames and 112 corresponding original frames.

Subjective quality evaluation experiments were conducted following the ITU-R BT.500-13\cite{bt2012500} to obtain opinion scores for the interpolated frames.   
A total of 21 inexperienced subjects participated in the subjective experiments, most of whom were college students from various disciplines. To simulate real-world conditions, subjects were allowed to evaluate the quality by either playing three frames consecutively or examining each frame individually. They were instructed to subjectively assess the quality of the interpolated frame and the transition between the triplet frames. Following convention\cite{9075375,8000398}, score ratings were divided into five levels as shown in Fig.~\ref{fig02a}.  Subjects rated the overall sensation of coherence and frame quality on a continuous quality scale from 0 to 100. After collecting subjective scores from interpolated frames, the mean opinion score (MOS) were computed. The distribution of MOSs for all interpolated frames and every VFI algorithms can be seen in Fig.~\ref{fig02b} and Fig.~\ref{fig02c}.

\section{The Proposed Method}
This section provides a detailed explanation of the proposed VFIPQA method. The triplet network is utilized in the method to extract features from succession of similar frames and the coherence similarity measurements are designed for quality assessment. 
\subsection{Feature Extract}
\begin{figure}[tb]
\centerline{\includegraphics[width=0.9\linewidth]{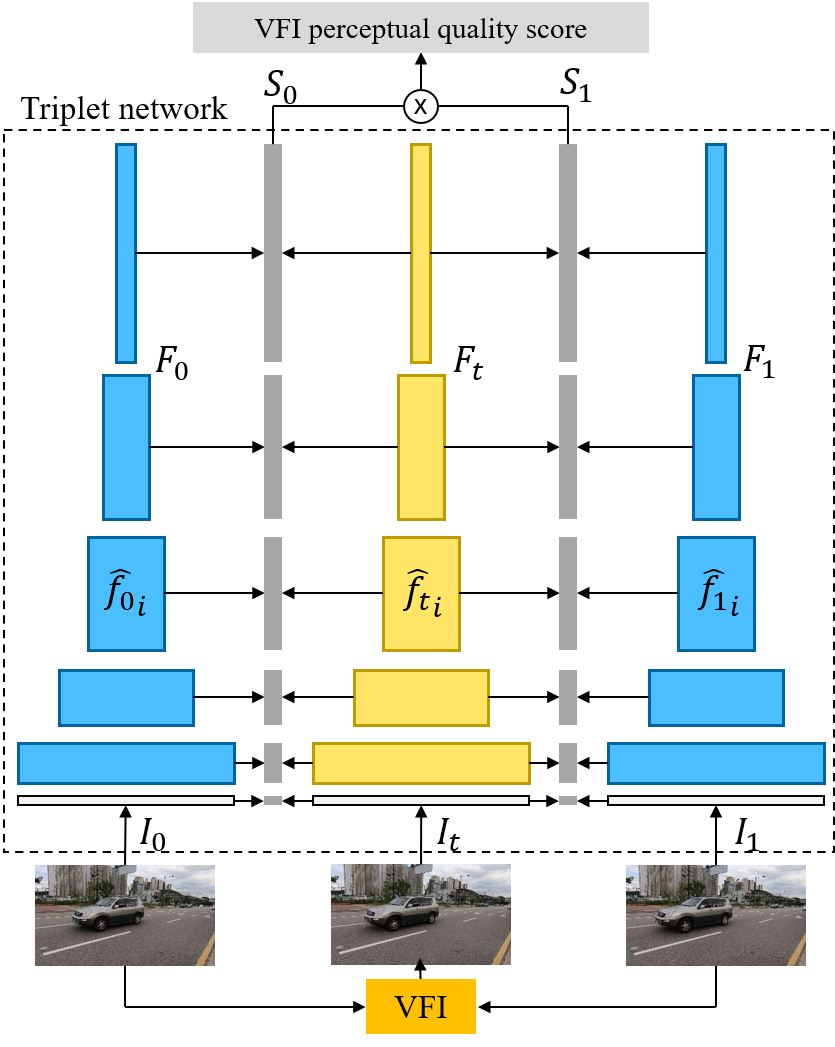}}
\caption{The network architecture of the proposed method. The network includes triplet feature networks to process three successive frames and the quality score of the interpolated frame is computed by coherence similarities.}
\label{fig1}
\end{figure}

Given that VFI involves the temporal interpolation of two frames into a new frame, which often results in a uncertain shift compared to the original frames.
And the human eye, when observing consecutive frames, will successively judge them based on the combination of the adjacent changes before and after.
On the basis of this prior, the framework of triplet network is introduced in feature extract.
The triplet network comprises three parallel pipelines, as illustrated in Fig.~\ref{fig1}. The inputs on sides are the original frames $\{I_0, I_1\}$, while the intermediate input $I_t$ is the interpolated frame from VFI. 

Convolutional neural networks (CNNs) have demonstrated strong capabilities in learning representations for human vision. 
Pre-trained CNN models have been found to align well with human subjective perception of image distortions\cite{zhang2018unreasonable}. 
Hence the ResNet\cite{he2016deep}, which exhibits strong representational capacity with fewer parameters, is adopted as the backbone in feature extract. 
The ResNet consists of five stages, each responsible for extracting features from different levels of an image, ranging from low to high. 
These stages of ResNet are as feature extractors to capture frame features at various scales and levels. Designate the feature maps of the $i$-th stage as $\hat{f_t}_i\in\mathbb{R}^{H_i\times W_i\times C_i}$, where $H_i,W_i,C_i$ is determined by the stages of ResNet.
To retain the maximum information, the input frames are preserved as features at the zeroth stage.
The whole feature maps are represented as:
$F_t=\{\hat{f_t}_i; i=0,\dots,5\}$.
The symbol $t$ stands for a interpolated image on a certain time , while $F_0$ and $F_1$ correspond the two original inputs of VFI. Introduced the original information, the extracted features are able to represent temporal coherence for similarity learning.

\subsection{Coherence Similarity}
To assess frame quality, the coherence similarity is calculated based on the deep features extracted from the CNNs. The feature maps obtained at different stages are find still retaining spatial structure features of the input image to varying degrees\cite{ding2020image}. Hence, it is reasonable to perform comparisons within the feature maps.
Both the structure similarity and texture similarity metrics\cite{wang2004image} are introduced to obtain coherence.
The similarities are calculated between the interpolated frame and the two input frames getting $(S_0, S_1)$ respectively.
In terms of $S_0$, the coherence similarity is defined as:
\begin{equation}
S_{0_i}^l=\frac{\mu^i_{0}\mu^i_{t}+c_1}{{\mu^i_{0}}^2+{\mu^i_{t}}^2+c_1},
S_{0_i}^s=\frac{\sigma^i_{0t}+c_2}{{\sigma^i_{0}}^2+{\sigma^i_{t}}^2+c_2}, \label{eq2}
\end{equation}
where $\mu^i_{0}$, $\mu^i_{t}$ are the global means in the stage $i$ of feature maps $F_0$ and $F_t$, $\sigma^i_{0}$, $\sigma^i_{t}$, and $\sigma^i_{0t}$ are the global variances and covariance respectively, and they keep the original channels $C_i$.
The $c_1$ and $c_2$ are the small positive constants to avoid numerical instability.
$S_1$ is calculated in the same way. Similarity computation on only one side will fail to learn coherence.

The metrics is utilized to predict the quality of the interpolated frames directly. Although VFI algorithms usually interpolate frames during intermediate moments, with short time intervals between frames. 
The relative motion of the same object in the interpolated frame and the input frames is not spatially symmetric absolutely. 
As a result,  $S_0$ and $S_1$ are assigned to different weights and multiplied together.
Learnable weights $\alpha_i$ and $\beta_i$ are introduced in each stage $i$ and sum together to form the final coherence similarity metric:
\begin{equation}
CS(S_0,S_1)=\sum_i{(\alpha_i S^l_{0_i} S^l_{1_i}+\beta_i S^s_{0_i} S^s_{1_i})}. \label{eq4}
\end{equation}
The $\alpha_i$ and $\beta_i$ are used to capture the sensitivity of deep features at different stages.
Calculated the coherence similarity between the interpolated frame and the input frames, a quality score for the interpolated frame can be obtain. 

\section{Experiments and Results}

\begin{table}[tb]
\caption{Performance comparison between state-of-art FR/NR IQA methods and the proposed method.}
\renewcommand\arraystretch{1.1}
\renewcommand\tabcolsep{10.0pt}
\begin{center}
\begin{tabular}{c ccc c}
\hline
\hline
\textbf{Method} & \textbf{SRCC}& \textbf{KRCC}& \textbf{PLCC} & \textbf{RMSE} \\
\hline
PSNR& 0.1305& 0.0854& 0.3098& 16.7217 \\
SSIM\cite{wang2004image}& 0.1715& 0.1196& 0.2925&16.7783  \\
FSIM\cite{zhang2011fsim}& 0.3350& 0.2347& 0.3829& 15.6374 \\
GMSD\cite{xue2013gradient}& 0.2448& 0.1687& 0.3422 & 16.3394 \\
LPIPS-Alex\cite{zhang2018unreasonable}& 0.6621& 0.4859& 0.6671 & 12.7870 \\
LPIPS-VGG\cite{zhang2018unreasonable}& 0.4877& 0.3437& 0.5381 & 14.6083 \\
DISTS\cite{ding2020image}& 0.7667& 0.5781& 0.8049& 10.3202 \\
\hline
BIQI\cite{moorthy2010two}& 0.4282& 0.2957& 0.4526& 15.5193 \\
BLIINDS-II\cite{saad2012blind}& 0.3373& 0.2391& 0.4331& 15.6733 \\
BRISQUE\cite{mittal2012no}& 0.3028& 0.2177& 0.4001& 15.9651 \\
DIIVINE\cite{moorthy2011blind}& 0.3941& 0.2708& 0.4138& 15.7943 \\
BMPRI\cite{min2018blind}& 0.3698& 0.2569& 0.4128& 15.8324 \\
NIQE\cite{mittal2012making}& 0.4507& 0.3176& 0.4962& 15.0690 \\
WaDIQaM\cite{bosse2017deep}& 0.6002& 0.4352&0.5611& 16.8733 \\
DBCNN\cite{zhang2018blind}& 0.7921& 0.6070&0.8058& 10.1555 \\
HyperIQA\cite{su2020blindly}& 0.7337& 0.5428& 0.7504& 11.4378 \\
MANIQA\cite{yang2022maniqa}& 0.7067& 0.5172&0.6941& 12.5253 \\
\hline
\textbf{Proposed}& \textbf{0.8248}& \textbf{0.6415} &\textbf{0.8197}& \textbf{9.9281} \\
\hline
\hline
\end{tabular}
\label{tab1}
\end{center}
\end{table}

\subsection{Implementation Details}

\begin{figure}[htb]
\subfloat{\includegraphics[width=0.24\textwidth]{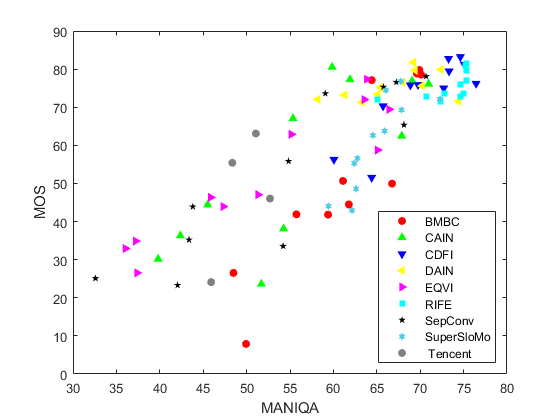}}
\subfloat{\includegraphics[width=0.24\textwidth]{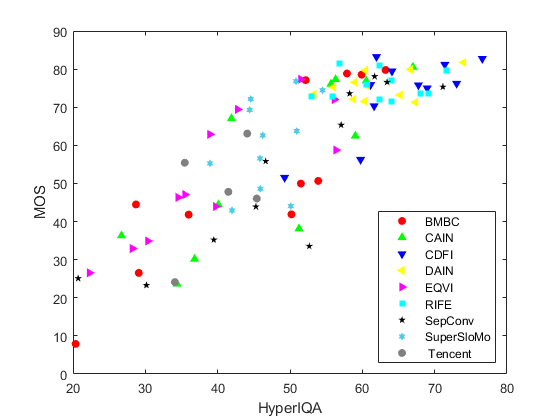}}
\\
\subfloat{\includegraphics[width=0.24\textwidth]{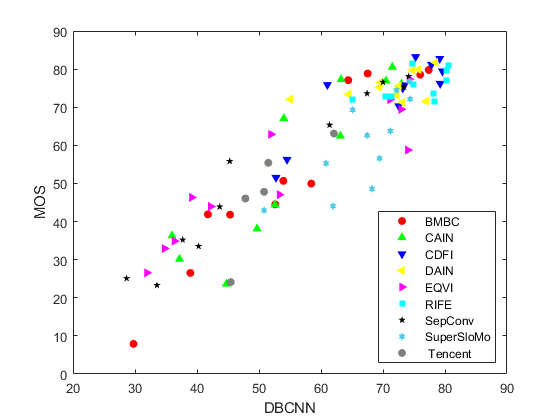}}
\subfloat{\includegraphics[width=0.24\textwidth]{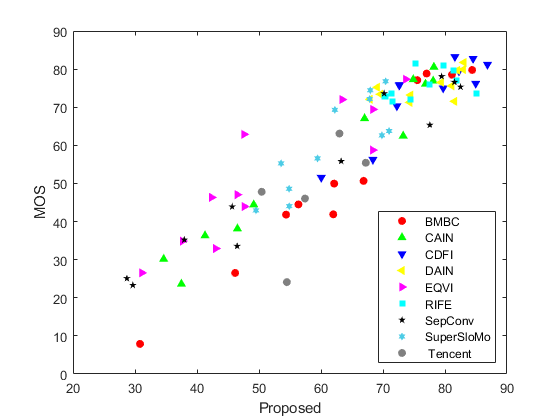}}
\caption{Scatter plots of the proposed method and three best-performing network based NR methods on the test dataset.}
\label{fig3}
\end{figure}

The proposed method is implemented using the PyTorch framework.
The ResNet50 is pre-trained on the ImageNet database.
The model parameters are optimized using the ADAM optimizer with an initial learning rate of $10^{-4}$, which is reduced by a factor of 2 after every 50 iterations.
The training process is performed on NVIDIA RTX 2080TI GPUs.
As convention\cite{8326697}, the dataset is split into two subsets: $80\%$ for training and $20\%$ for testing, ensuring there is no overlap between the two sets. The training loss is the Mean Square Error (MSE) loss. To ensure robustness and reliability, the training and testing process are repeat 10 times at different split data and report the average values of the evaluation criteria as the final results. 

\subsection{Experimental Results}
Six FR IQA algorithms and eleven NR IQA algorithms are selected for performance comparison.
The compared FR IQA algorithms include traditional methods\cite{wang2004image,zhang2011fsim, xue2013gradient} and network based methods\cite{zhang2018unreasonable, ding2020image}.
The compared NR IQA algorithms also include traditional methods \cite{moorthy2010two, moorthy2011blind, mittal2012no, saad2012blind, mittal2012making, min2018blind} and network based methods\cite{bosse2017deep, zhang2018blind, su2020blindly, yang2022maniqa}.
For evaluating the performance of the algorithms, four commonly used performance criteria: the Spearman Rank Order Correlation Coefficient (SRCC), Kendall Rank-Order Correlation Coefficient (KRCC), Pearson Linear Correlation Coefficient (PLCC), and Root Mean Square Error (RMSE) are employed as evaluation metrics \cite{seshadrinathan2010study}. Correlation coefficient is used to assess the consistency of the objective method scores with the MOS.
The experimental results on the new dataset are presented in Table~\ref{tab1}.
The FR methods are applied on their well-trained perceptual model directly as it is used in VFI as evaluation metrics.
All NR methods are retrained following their respective training protocols.

The results demonstrate that the proposed method achieves significantly higher values compared to all the other algorithms. The scatter plots shown as in Fig.~\ref{fig3} indicate that the proposed method is more clustered than NR network methods, which implies better consistency with MOS..
It is worth noting that both traditional and learning-based NR IQA methods show a decrease in performance when applied to VFI quality assessment, as compared to their performance on generic IQA databases\cite{bosse2017deep, zhang2018blind, su2020blindly, yang2022maniqa}.
This observation highlights the specificity of perceptual quality assessment for VFI and emphasizes the importance of constructing a dedicated VFI quality assessment dataset.
Furthermore, in comparison to FR IQA algorithms, the proposed method does not rely on intermediate reference frames for evaluation. This characteristic enhances the applicability and versatility of the method in various VFI scenarios.

Ablation experiments are provided to illustrate the effect of the backbones of feature extraction and the coherence similarity, as in Table~\ref{tab2}.
The proposed method uses ResNet50 as a backbone outperforming the Swin Transformer in feature extraction.
It may be that ResNet is easier to learn the coherent features.
In addition, the designed coherence similarity is verified by comparison of the single similarity. This suggests that the learning of coherent information is necessary to include both front and back inputs.

\begin{table}[tb]
\caption{Ablation studys on the proposed method.}
\renewcommand\arraystretch{1.1}
\renewcommand\tabcolsep{10.0pt}
\begin{center}
\begin{tabular}{cccc}
\hline
\hline
\textbf{Components} & \textbf{SRCC}& \textbf{PLCC}& \textbf{RMSE} \\
\hline
VGG16 \cite{simonyan2014very}& 0.7945& 0.7925 & 10.6353  \\
Swin \cite{liu2021swin}& 0.8008& 0.8068 & 10.2986\\
AlexNet \cite{krizhevsky2012imagenet}& 0.7895& 0.7849 & 10.8944  \\
ResNet50 (single)& 0.7502 & 0.7455 & 11.5988   \\
\textbf{Proposed}& \textbf{0.8248} & \textbf{0.8197} & \textbf{9.9281}  \\
\hline
\hline
\end{tabular}
\label{tab2}
\end{center}
\end{table}

\section{Conclusion}
This paper presents a comprehensive approach for VFI perceptual quality assessment. The proposed method extracts features from the interpolated and input frames of VFI and the perceptual quality of the interpolated frame is then computed using the designed coherence similarities. Subject experiments on the VFI frames, consisting of frames interpolated by various VFI algorithms, are constructed to obtain a quality dataset.
Experimental results on the new dataset validate the effectiveness of the proposed method in assessing the perceptual quality of VFI. This work will help VFI research to move further towards high quality frames.

\bibliographystyle{IEEEtran}
\bibliography{references}{}

\end{document}